\documentclass[a4paper,11pt]{article}
\pdfoutput=1 % if your are submitting a pdflatex (i.e. if you have
\usepackage{jcappub} % for details on the use of the package, please
\usepackage[T1]{fontenc} % if needed

\title{\boldmath Relating Braking Indices of Young Pulsars to the Dynamics of Superfluid Cores}

\author[a,1]{H. O. Oliveira,\note{Corresponding Author}}
\author[a,b]{N. S. Magalhaes,}
\author[a]{R. M. Marinho, Jr.,}
\author[a,c,d]{G. A. Carvalho}
\author[e]{and C. Frajuca}
\affiliation[a]{Graduate Program in Physics, Technological Institute of Aeronautics, 
             Pra\c{c}a Marechal Eduardo Gomes 50, Sao Jose dos Campos, SP 12228-900, Brazil}
\affiliation[b]{Federal University of Sao Paulo, Department of Physics, 
  Rua Sao Nicolau 210, Diadema, SP  09913-030, Brazil}
\affiliation[c]{Dipartimento di Fisica and ICRA, Sapienza Universit\`a di Roma, P.le Aldo Moro 5, I--00185 Rome, Italy.}
\affiliation[d]{ICRANet, P.zza della Repubblica 10, I--65122 Pescara, Italy.}
\affiliation[e]{Federal Institute of Education, Science and Technology of Sao Paulo, R. Pedro Vicente 625, Sao Paulo, SP 01109-010, Brazil}

\emailAdd{heitoroliveiradeoliveira@gmail.com}
\abstract{Pulsars are stars that emit electromagnetic radiation in well-defined time intervals. The frequency of such pulses decays with time as is quantified by the {\it braking index} ($n$). In the canonical model $n = 3$ for all pulsars, but observational data show that $n \neq 3$, indicating a limitation of the model. In this work we present a new approach to study the frequency decay of the rotation of a pulsar, based on an  adaptation of the canonical one. We consider the pulsar a star that rotates in vacuum and has a strong magnetic field but, differently from the canonical model, we assume that  its moment of inertia changes in time due to a uniform variation of a displacement parameter in time. We found that the braking index results smaller than the canonical value as a consequence of an increase in the star's displacement parameter, whose variation is small enough to allow plausible physical considerations that can be applied to a more complex model for pulsars in the future. In particular, this variation is of the order of neutron vortices' creep in rotating superfluids. When applied to pulsar data our model yielded values for the stars' braking indices close to the observational ones. The application of this approach to a more complex star model, where pulsars are assumed to have superfluid interiors, is the next step in probing it. We hypothesize that the slow expansion of the displacement parameter might mimic the motion of core superfluid neutron vortices in realistic models.}

\begin{document}
\maketitle
\flushbottom

\section{Introduction} \label{sec:intro}

Pulsars are astrophysical objects normally modeled as neutron stars that originated from the collapse of a progenitor star \cite{wu2003}. In a model which we will refer to as canonical, pulsars are described by spherical magnetized dipoles that rotate, usually with the magnetic axis misaligned relatively to the rotation axis. This misalignment would be responsible for the observation of radiation emitted in well-defined time intervals in a certain direction, which is the typical observational characteristic of this kind of star.

Observations show that the rotation frequency of pulsars is slowly decaying with time (the stars are spinning down), implying a gradual decrease of the angular velocity, according to the canonical model \cite{Chukwude2010}. Such decay can be quantified by a dimensionless parameter known as {\it braking index}, $n$. The canonical model predicts that this index has only one value for all pulsars, equal to 3. However, observational data show that actual braking indices are different from 3, indicating that the canonical model requires improvement \cite{Chen2006}.

One such attempt of improvement is the pulsar wind model \cite{xu01}, which yields a braking index n=1 but has presented interesting results when combined with the canonical contribution due to the magnetic dipole (n=3). A similar reasoning is followed in the work \cite{kou15}.

Other works investigated different aspects of the problem. For instance, \cite{magalhaes12} proposed a phenomenological function in the energy conservation formula, while \cite{allen97} proposed an increase in the angle between the magnetic moment and rotation axis as the cause of the evolution of the torque that has been related to glitch events for non-canonical braking indices. 
Moreover, \cite{magalhaes16} investigated the possibility of an effective force acting on the star, whereas \cite{gao17} and \cite{Chen2016} studied the high braking index of PSR J1640-4631 combining magneto-dipole radiation and dipole magnetic field decay. In addition, \cite{kou2014} presented the possibility that accretion of a millicharged dark matter can change the braking index of a pulsar.

In order to explain braking indices we model a pulsar as an object composed by a large superfluid core and a thin outer crust, which may couple and decouple according to physical conditions. Such coupling and decoupling dynamics may introduce sudden changes in rotation as well as variations on the effective moment of inertia in the long-term, yielding changes to the canonical model \cite{ho12}. 

The main result of this work is to revisit the canonical model introducing a {\it displacement parameter} which we show (in section \ref{sec:3:A}) to resemble the radial velocity of the flow of superfluid vortices of the core. Starting from the model of spin-powered pulsars, where the electromagnetic energy is provided by the rotational energy \cite{gold68}, we explore the influence of the variation of the displacement parameter on the value of the braking index considering a theoretical model. According to our approach when the braking index is less than $3$ the vortices would influence the star's angular momentum, increasing it without changing the emission of magnetic dipole energy. The increase of the displacement parameter would reduce the star's angular velocity and, consequently, the frequency of the light pulses of the pulsar, as is observed.

We tested our approach with data of ten pulsars with observed braking indices, obtaining theoretical indices compatible with the observational ones. Also, we found that the variation of the displacement parameter is small and slow  in time, suggesting that the related physical  process is feasible.

This paper is organized as follows. In section \ref{sec:2} we briefly review the canonical model as also a description about superfluid core and list the ten pulsar sample. In section \ref{sec:3} we present our pulsar model, which bears great similarity to the canonical one except for the radial velocity flow. In section \ref{sec:4} we discuss physical mechanisms. Building on the knowledge developed with the ten pulsars, in section \ref{sec:5} we predict braking indices for additional young pulsars.  Section \ref{sec:conc} closes the paper with our concluding remarks.

%#################################################################

\section{Summary of the canonical model and observed braking indices} \label{sec:2}

In the model usually adopted to describe pulsars, to which we refer as canonical \cite{pacini67, gold68}, these stars are composed mainly by neutrons and are described by spherical magnetized dipoles that rotate with angular velocity $\Omega = 2 \pi \nu$, with the magnetic axis misaligned relatively to the rotation axis. The electromagnetic energy radiating from the pulsar is given by \cite{griffiths99, shapiro08},
\begin{equation}
\dot{E}_{mr}=\frac{2}{3c^3}|\ddot{m}|^2,
\label{eq:dotEmr}
\end{equation}
where the dot denotes a time derivative, $c$ is the speed of light in vacuum and  $\vec{m}$ is the dipole moment,
\begin{equation}
\vec{m}=\frac{B_PR_0^3}{2}(\cos \alpha \hat{k}+\sin \alpha \cos (\Omega t) \hat{i}+\sin\alpha \sin(\Omega t) \hat{j}).
\label{eq:magnMom}
\end{equation}
This expression has the following {\it constants:} $B_P$ is the magnetic field at the pole, R$_0$ is the radius of the pulsar and $\alpha$ is the angle between the magnetic dipole axis and the rotation axis. 

The rotation frequency, $\nu$, of the pulsar varies with time, as shown in Table \ref{table:01}. This could reflect on the behavior of the magnetic field with time since in this model \cite{glend96} it has the following expression when $\sin\alpha = $ 1  (which is the case when the dipole moment is perpendicular to the rotation axis): 
\begin{equation}
B_P= \sqrt{\frac{12 c^3 M}{5 R_0^4}} \sqrt{\frac{- \dot \Omega}{\Omega ^3}}.
\label{eq:B_P_can}
\end{equation}
However, since the rotation frequency and its derivatives vary very slowly with time, for practical purposes B$_P$ is normally considered constant. We use this expression to obtain the values of ``$B_P$ (canon)'' in Table \ref{table:02}.

Using equations (\ref{eq:magnMom}) and (\ref{eq:B_P_can}) the expression for the radiated energy becomes
\begin{equation}
\dot{E}_{mr}=\frac{1}{6c^3}B_P^2R_0^6\Omega^4 \sin ^2\alpha.
\label{eq:derivadaenergiamag}
\end{equation}

In the canonical model the energy carried away by the radiation originates exclusively from the rotational kinetic energy of the neutron star, given by
\begin{equation}
{E_{rot}=\frac{1}{2}I\Omega^2},
\label{eq:energiarot}
\end{equation}
where $I$ is moment of inertia of a sphere, assumed constant. (For estimate purposes in some works the moment of inertia is calculated using the approximate formula $I = MR_0^2$. Here we use $I= 2 M R_0^2/5$.) The rotation power is thus
\begin{equation}
\dot{E}_{rot}= I\cdot \Omega \cdot \dot{\Omega}.
\label{eq:Erot}
\end{equation}

Energy conservation translated in terms of power yields
\begin{equation}
\dot{E}_{rot}=-\dot{E}_{mr},
\label{eq:VEtotal}
\end{equation}
which in this model implies
\begin{equation}
\dot{\Omega}=k\Omega^3.
\label{spindown}
\end{equation}
Therefore, the canonical model implies a gradual slowdown of the star's rotation.

A pulsar's spin-down can be quantified by the braking index, $n$, defined by
\begin{equation}
n \equiv \frac{\Omega \ddot{\Omega}}{\dot{\Omega}^2}.
\label{eq:article3}
\end{equation}
For the canonical model equation (\ref{spindown}) yields $n=3$, a value constant for all pulsars. However, observational data have resulted in $n \neq 3$.

For the pulsars listed in Table \ref{table:01} data is not dominated by timing noise;  in the case of the Vela pulsar, a more accurate method was used to obtain data in the long term \cite{espinoza16}. Therefore, reliable calculations of braking indices are possible for such pulsars, and they yield braking indices smaller than the value given by the canonical model. This inconsistency is the motivation for the approach that we propose in this work.

%#################################################################
   \begin{table*}
   	\caption{\label{table:01} Rotation frequency ($\nu$) and its first and second time derivatives for the sample of pulsars.}
  	\centering   	
        \small{
        \begin{tabular}{ccccccc}
   	\hline
   	\noalign{\smallskip}
Name & J name & $\nu$  & $\dot{\nu}$ & $\ddot{\nu}$ & n & Refs.\\
	&     & (s$^{-1}$)  & ($\times 10^{-10}$ s$^{-2}$) & ($\times 10^{-21}$ s$^{-3}$)  & &\\
   	\noalign{\smallskip}
   	\hline
   	\noalign{\smallskip}
B 0531+21	& J0534+2200 & 29.946923		& -3.77535		& 11.147	& 2.342(1)	& \cite{lyne15}\\
%J0537-6910	& J0537-6910 & 62.018715011		& -1.99324		& -0.77		& -1.2	& \cite{antonopoulou18}\\
B 0540-69	& J0540-6919 & 19.7746860321	& -1.8727175	& 3.772		& 2.13(1)	& \cite{ferdman15}\\
B0833-45	& J0835-4510 & 11.200			& -0.15375		& 0.036		& 1.7(2)	& \cite{espinoza16}\\
J1119-6127	& J1119-6127 & 2.4512027814 	& -0.2415507	& 0.6389 	& 2.684(2)$^a$	& \cite{weltevrede11}\\
J1208-6238	& J1208-6238 & 2.26968010518	& -1.6842733	& 0.33		& 2.598(1)	& \cite{clark16}\\
B1509-58	& J1513-5908 & 6.611515243850	& -0.6694371307	& 1.9185594	& 2.832(3)	& \cite{livingstone11}\\
J1734-3333	& J1734-3333 & 0.855182765		& -0.0166702	& 0.0028	& 0.9(2)	& \cite{espinoza11}\\
J1833-1034	& J1833-1034 & 16.15935711336	& -0.52751130	& 0.3197	& 1.857(1)	& \cite{roy12}\\
J1846-0258	& J1846-0258 & 3.059040903		& -0.665131		& 3.17		& 2.64(1)$^b$ & \cite{livingstone11}\cite{livingstone07}\cite{archibald15}\\
   	\noalign{\smallskip}
   	\hline
   	\end{tabular}}
    \\	Notes. Besides these references, information regarding associations and most rotational parameters were taken from the ATNF Pulsar catalogue (http://www.atnf.csiro.au/research/pulsar/psrcat/; Manchester et al. 2005). Uncertainties ($1\sigma$) on the last quoted digit are shown between parentheses.\\
    $^a$ A possible reduction of about 15\% is observed after a large glitch \cite{antonopoulou15}.\\
    $^b$ The braking index was found to decrease to $n=2.19$ after a large glitch \cite{livingstone11b,archibald15}.
    \end{table*}
%#################################################################

%%%%%%%%%%%%%%%%%%%%%%%%%%%%%%%%%%%%%

\section{State of Thermodynamic Equilibrium of the Core of a Rotating Neutron Star}\label{sec:3:A}

In order to propose a refinement to the canonical model we first focus on a more complex model for the neutron star, starting from its birth.

In the initial stage of the neutron star after the explosion of its progenitor star (supernova effect), matter is in a normal state due to the temperatures reached after the collapse of the star core. The star quickly cools down through neutrino irradiation \cite{manchester77} and when the temperature is below Cooper's critical temperature neutrons form a superfluid \cite{migdal60} while protons form a superconductor \cite{baym69}. This critical temperature is given by the ratio between the Fermi energy and the Boltzmann constant ($Tc=\frac{E_F}{k_B}$), and for neutron matter it is close to $10^{12}$ K which is much greater than the value of the inner temperature of neutron stars ($10^6-10^8$ K) \cite{sauls89}. Therefore, we admit that neutron stars have a thin, solid outer crust that surrounds the majority of its mass, which lays in the core mostly in the form of superfluid neutrons and superconducting protons.

As the crust rotates the superfluid is expected to move as well, reaching a dynamical state dominated by vortices whose axes are parallel to the rotation axis. The influence of these vortices is quantified through a number known as  circulation. We brefiely summarize the general theory as follows, for the sake of completeness.

Given a vortex in a neutron superfluid, the {\it circulation} is defined as the line integral  along a path $C$ that surrounds the vortex circulation, given as \cite{ghosh07,landau80}):
\begin{equation}
\kappa = \oint _C \vec{v}_s\cdot d\vec{l},\label{cap:04:eq:01}
\end{equation}
where $v_s$ is the fluid velocity and $dl$ the line element along $C$. 

The fluid velocity can be found considering that the general shape of the Cooper pair can be described by a wave function of the condensed as \cite{sauls89,ghosh07}
\begin{equation}
\psi=|\psi_a|e^{i\theta},\label{cap:04:part2:eq:02}
\end{equation}
where the amplitude $|\psi_a|$ is a thermodynamic state variable and the phase $\theta$ is a scalar. 

From this wave function the velocity of the superfluid is given by
\begin{equation}
\vec{v_s}=\frac{\hbar}{2m_n}\vec{\nabla}\theta,\label{cap:04:eq:02}
\end{equation}
with $2m_n$ being the mass of a pair of neutrons and $\hbar$ being Planck's constant divided by 2$\pi$ . As this velocity is the gradient of a scalar it follows that the flow of a superfluid is irrotational, implying that the condensate will be able to withstand circulation only at certain points of singularities within the fluid \cite{sauls89}. In such isolated configuration with non-zero circulation the singularities are vortex lines \cite{ghosh07}, which carry angular momentum. The greater (smaller) the angular momentum of the superfluid, the more (less) vortex lines it will harbor.

Vortex conservation law is associated with mass conservation equation in fluid mechanics \cite{ghosh07,gobbi06}. When this law is associated to other superfluid properties \cite{alpar84,alpar89} one finds an expression for the brake in the superfluid rotation, $\dot{\Omega}_s$:

\begin{equation}
\dot{\Omega}_s=-\frac{\kappa_0n_vv_R}{R_n},\label{cap:04:eq:37}
\end{equation}
where $v_R$ is the radial velocity (relative to the rotation axis of the star) of the {it flow of the superfluid vortices}, $R_n$ is the radius of the star's core  and $n_v$ is the density of vortices. The quantum of circulation, $\kappa_0$, is given by $\kappa_0 = \pi \hbar/m_n$. Therefore, when the superfluid rotation decays in time  the vortices move outward with velocity $v_R$.

One can show that \cite{hall60} $n_v\kappa_0=2\Omega_s$, allowing eqn. (\ref{cap:04:eq:37}) to be rewritten as
\begin{equation}
\frac{\dot{\Omega}_s}{\Omega_s}=-\frac{2 v_R}{R_n},\label{cap:04:eq:37_a}
\end{equation}

After a pulsar is rotating for a while, core and crust are expected to rotate at the same angular velocity (except when special, short-term events might occur, like glitches). Therefore, in the long term we have $\Omega_s = \Omega$. For the same reason it results that as the crust brakes due to magnetic torque the core brakes correspondingly, implying $\dot{\Omega}_s\equiv\dot{\Omega}$. In this case eqn. (\ref{cap:04:eq:37_a}) is given by
\begin{equation}
\frac{\dot{\Omega}}{\Omega}=-\frac{2 v_R}{R_n},\label{cap:04:eq:37_b}
\end{equation}

It is estimated that the crust of a pulsar is thin, occupying approximately 10\% of its radius. Therefore, considering a typical total radius of 10 km the core's radius is  $R_n \approx$ 9 km. Given this figure, typical vortices' velocities can be estimated from observational data of pulsars' rotation frequencies. For example, using data from Table \ref{table:01} for the Crab pulsar (B 0531+21) we obtain $v_R \approx$ 0.5 cm/day, showing that Crab's superfluid vortices move quite slowly away from the rotation axis.

This kind of estimate is important to help probing the inner structure of neutron stars as the number obtained is connected to observational data. As \cite{alpar89} quoted, each physical model for superfluid spin decay is in general a model for $v_R$. In the next section we will approach this result from a different perspective, revisiting the canonical model and relating the core superfluid dynamics to the braking index of pulsars.

%##############################################################

\section{Braking indices of Pulsars with Dynamic Cores}\label{sec:3}

\subsection{The Displacement Parameter and Rotation Power}\label{subsec:MDRDP}

Pulsars may be treated as slowly rotating stars with moments of inertia very close to those of non-rotating neutron stars \cite{ostriker69}. \cite{hartleThorne1968} showed that rigid and slow rotation distorts neutron stars, changing their moments of inertia; it also promotes a spherical stretching that varies with the star's radius, which can be quantified by the fractional change in mean radius, depending on the chosen equation of state. In general rotation tends to increase a pulsar's radius relative to the non-rotating regime.

This stretching of a pulsar in response to rotation inspired this investigation of the behavior of the braking index in the presence of a variation in time of a convenient displacement parameter.

As in the canonical model, we assume that the pulsar changes its rotational energy ($E_{rot}$) into the form of electromagnetic dipole radiation (${E}_{mr}$) according to equation (\ref{eq:VEtotal}). However, differently from that model we will consider that the pulsar consists of a thin, solid crust with {\it constant} moment of inertia I$_c$, and a large spherical core with total constant mass, $M_n$, made basically of superfluid neutrons, whose moment of inertia will be given by
\begin{equation}
I_n (t) \equiv \lambda M_n R^2(t). 
\label{eq:momInCore}
\end{equation}

In this expression $R$ is a {\it displacement parameter} that summarizes in its mathematical behavior all physical factors that influence the moment of inertia other than the core's total mass. We  assume that the core's moment of inertia may change with time but not due to changes in its total mass ($M_n$) or physical radius ($R_n$); instead, any change in $I_n$ should be due to {\it internal displacements} of mass that are quantified by $R(t)$. 

It is natural to expect that the displacement parameter will be close to $R_n$, so we perform a Taylor expansion of this parameter as follows:
\[
 R(t)= R_n +\dot{R}(t_0)\cdot (t-t_0)+\ddot{R}(t_0)\frac{(t-t_0)^2}{2}+...
\]
where  $t_0$ is an instant in time that we will choose equal to zero.

We will assume that this expansion can be truncated after its second term due to negligible higher $R$ derivatives:
\begin{equation}
R(t) \approx R_n + t \dot{R}.
\label{eq:rft1}
\end{equation}

As a first order approximation we will assume that the first time derivative of the displacement parameter ($\dot R$) is constant. This number, with units
 of velocity, is expected to vary from pulsar to pulsar as it informs about the inner dynamics of the star. 

Note that when $\dot R = 0$ equation (\ref{eq:rft1}) yields $R = R_n$ and equation (\ref{eq:momInCore}) results in the usual moment of inertia of a solid core as expected. On the other hand, a nonzero $\dot R$ in equation (\ref{eq:rft1}) does not mean a change in the value of the physical radius of the core; it indicates the presence of internal displacements of mass that change the core's moment of inertia, as explained above.  

Differentiating the expression for the rotational energy, equation (\ref{eq:energiarot}), with respect to time, in view of the assumptions of our model results in
\begin{equation}
\dot{E}_{rot}= \frac{1}{2} \frac{d}{dt} ( I_c \Omega_c^2 + I_n \Omega_n^2).
\label{eq:article1a}
\end{equation}

Since any changes in the angular velocity of the crust, $\Omega_c$, are rapidly transmitted to the core, in practice the angular velocity of the latter, $\Omega_n$, will be considered equal to $\Omega_c \equiv \Omega$. Therefore equation (\ref{eq:article1a}) becomes
\begin{equation}
\dot{E}_{rot}= \frac{1}{2} \left[ ( I_c + I_n ) 2 \Omega \dot\Omega + \dot I_n \Omega^2 \right].
\label{eq:article1b}
\end{equation}

As the moment of inertia of the thin crust is expected to be much less than the moment of inertia of the large, heavy core, we approximate 
$I_c + I_n \approx I_n$. Similarly, we consider $M_n$ practically equal to the total mass of the pulsar, $M$. Using these approximations in equation  (\ref{eq:article1b}) together with the time derivative of the definition  (\ref{eq:momInCore}) we find
\begin{equation}
\dot{E}_{rot} = I_n \Omega \dot\Omega + \frac{2}{5} M R \dot R \Omega^2,
\label{eq:article1c}
\end{equation}
which, using the definition (\ref{eq:momInCore}), becomes
\begin{equation}
\dot{E}_{rot}=\lambda M R^2 \Omega^2 \left(
 \frac{\dot{\Omega}}{\Omega} + \frac{\dot{R}}{R} \right).
\label{eq:article1}
\end{equation}

We will further admit that for the duration of typical observational time intervals, $\tau$, the condition $\tau \dot{R} << R_n$ holds such that equation (\ref{eq:rft1}) yields the following typical value for $R$:
\begin{equation}
R = R_n \left( 1 + \tau \frac{ \dot{R}}{R_n} \right) \Rightarrow R \approx R_n .
\label{eq:rft1a}
\end{equation}
Therefore, for typical observational times equation (\ref{eq:article1}) assumes the form
\begin{equation}
\dot{E}_{rot}=\lambda M R_n^2 \Omega^2 \left(
 \frac{\dot{\Omega}}{\Omega} + \frac{\dot{R}}{R_n} \right).
\label{eq:article1new}
\end{equation}

Finally, assuming a core that occupies the vast majority of the pulsar's volume we have $R_n \approx R_0$ and the expression for the rotation power becomes
\begin{equation}
\dot{E}_{rot}=\lambda M R_0^2 \Omega^2 \left(
 \frac{\dot{\Omega}}{\Omega} + \frac{\dot{R}}{R_0} \right).
\label{eq:article1new1}
\end{equation}

%#####################################
\subsection{The Displacement Parameter and Braking Indices}\label{subsec:DPBI}

Our assumptions can also be used to find the expression for the magnetic radiation power from equations (\ref{eq:dotEmr}) and (\ref{eq:magnMom}), which results in
\[
\dot{E}_{mr}=\frac{\sin ^2 \alpha B_P^2 R_0^6\Omega ^4+24 \sin ^2 \alpha B_P^2R_0^4\dot{R}^2\Omega ^2}{6c^3}+
\]
\begin{equation}
\frac{36B_P^2R_0^2\dot{R}^4}{6c^3}.
\label{eq:article2}
\end{equation}

Substituting equations (\ref{eq:article1new1}) and (\ref{eq:article2}) in equation (\ref{eq:VEtotal}) yields
\begin{eqnarray}
\nonumber \lambda M R^2 \Omega^2 \left(
 \frac{\dot{\Omega}}{\Omega} + \frac{\dot{R}}{R} \right)= -\frac{\sin ^2 \alpha B_P^2R^6\Omega ^4}{6c^3}\\
-\frac{24 \sin ^2 \alpha B_P^2R^4\dot{R}^2\Omega ^2+36B_P^2R^2\dot{R}^4}{6c^3},
\label{eq:314}
\end{eqnarray}
where we dropped the sub index $0$ in $R_0$ for simplicity such that $R$ henceforth corresponds to the typical star radius.

Solving this equation for the time variation of the angular velocity, $\dot\Omega$, we obtain an expression that we differentiate with respect to time to determine a new expression, which allows us to determine $\ddot{\Omega}$. Finally, we obtain the braking index \textit{n} using the definition (\ref{eq:article3}):
\begin{equation}
n=\frac{(3 \sin ^2\alpha B_P^2R^{5}\Omega ^2)}{(12 \lambda c^3 \dot{R} M+\sin ^2 \alpha B_P^2R^5 \Omega ^2)}.
\label{eq:braking}
\end{equation}

Solving equation (\ref{eq:braking}) for $\dot{R}$ we find the expression for the time variation of the displacement parameter:
\begin{equation}
\dot{R}=-\frac{\sin ^2 \alpha B_P^2(n-3)R^5\Omega ^2}{12 \lambda c^3 n M}.
\label{eq:raiovariando}
\end{equation}

We will estimate values for $\dot{R}$ assuming $c=2.9\times 10^{5}$ km s$^{-1}$ for the speed of light in vacuum and the following typical values applied to the pulsars given in Table \ref{table:01}: star radius R = 10 km and  total mass M $= 1.4$ M$_{\odot}$ (where M$_{\odot}$ denotes one solar mass). These values correspond closely to stable neutron stars as described in Ref. \cite{hartleThorne1968}. They imply,  when the pulsars were born, an estimated moment of inertia of I$_0$ = 2MR$^2$/5 = 56 M$_\odot$ km$^2$.

%#####################################
\subsection{The Displacement Parameter and Pulsars' Magnetic Fields}\label{subsec:DPPMF}

As for the magnetic field, it is common to adopt the typical value of $B_P=10^{12}$ G obtained from the canonical model. However, we will verify whether the time variation of the displacement parameter would change the magnetic field significantly in the present model. To this end we isolate the magnetic field in  equation (\ref{eq:314}), obtaining the expression 
\begin{equation}
B_P=\frac{\sqrt{6}c\sqrt{-\lambda c \Omega \dot{\Omega} M -\frac{\lambda c \dot{R} \Omega ^2 M}{R}}}{\sqrt{\sin ^2 (\alpha) R^4 \Omega ^4 +24 \sin ^2(\alpha) R^2 (\dot{R})^2 \Omega ^2 + 36(\dot{R})^4}}
\label{eq:Bp}
\end{equation}

In the above equation the second and third terms under the square root in the denominator are negligible when compared to its first term. Therefore, the expression for the magnetic field for pulsars in our model is approximated by
\begin{equation}
B_P= \sqrt{\frac{6\lambda c^3 M}{ R^4 sin^2 (\alpha) }} \sqrt{\frac{- \dot \Omega}{\Omega ^3}- \frac{\dot R}{R \Omega ^2}}.
%B_P=\frac{2\sqrt{3}c\sqrt{-c\Omega\dot{\Omega}M-\frac{c\dot{R}\Omega ^2 M}{R}}}{\sqrt{5}\sqrt{\sin ^2 (\alpha)R^4\Omega ^4}}.
\label{eq:Bp2}
\end{equation}
In this expression, when $\dot R=0$ and $\sin \alpha =1$ the canonical expression (\ref{eq:B_P_can}) is recovered.

The second term under the second square root of this equation will be negligible when
\begin{equation}
|\dot{R}| \ll |\frac{\dot{\Omega}}{\Omega}R|.
\label{eq:condicao}
\end{equation}
We will find to which extent this relation is satisfied after estimating $\dot R$ with the aid of the braking indices from Table \ref{table:01} as follows.

%#################################################################
\begin{table*}
	\caption{\label{table:02} Time variation of the displacement parameter ($\dot{R}$), magnetic field at the pole according to our model ($B_P$), magnetic field at the pole according to the canonical model ($B_P$ (canon.)), characteristic time ($t_c$) and moment of inertia after the characteristic time has elapsed ($I_{char}$).}
	\centering	
	\begin{tabular}{ccccccc}
		\hline
		\noalign{\smallskip}
Pulsar & $\dot{R}$  & $\frac{-\dot{\Omega}}{\Omega}R$ & $B_P$ & $B_P$ (canon.) &$t_c$ & $I_{char}$\\
& (cm s$^{-1}$) & (cm s$^{-1}$) & ($G$) & ($G$) & kys. & M$_\odot$ km$^2$\\
		\noalign{\smallskip}
		\hline
		\noalign{\smallskip}
B0531+21	& $1.6\times 10^{-6}$	& $1.3\times 10^{-5}$ 
			& $7.1\times 10^{12}$	& $7.6\times 10^{12}$	
			& $1.26$ & $63.4$\\
%J0537-6910	& \textcolor{red}{$7.51\times 10^{-6}$}	& \textcolor{red}{$3.21\times 10^{-6}$}
%			& \textcolor{red}{$2.15(-1)^{0.5}\times 10^{12}$} & \textcolor{red}{$1.86\times 10^{12}$} 
%			& \textcolor{red}{$4.93$}  &\\
B0540-69	& $1.6\times 10^{-6}$	& $9.5\times 10^{-6}$
			& $9.1\times 10^{12}$	& $1.0\times 10^{13}$	
			& $1.67$ & $65.8$\\
B0833-45	& $3.8\times 10^{-7}$	& $1.4\times 10^{-6}$   
			& $5.7\times 10^{12}$	& $6.7\times 10^{12}$	
			& $11.5$ & $72.5$\\
J1119-6127	& $5.5\times 10^{-7}$	& $9.9\times 10^{-6}$ 
			& $8.0\times 10^{13}$	& $8.2\times 10^{13}$ 	
			& $1.61$ & $59.2$\\
J1208-6238	& $4.7\times 10^{-7}$	& $7.4\times 10^{-6}$
			& $7.5\times 10^{13}$	& $7.7\times 10^{13}$	
			& $2.14$ & $59.6$\\
B1509-58	& $2.9\times 10^{-7}$	& $1.0\times 10^{-5}$ 
			& $3.0\times 10^{13}$	& $3.1\times 10^{13}$	
			& $1.57$ & $57.6$\\			
J1734-3333	& $1.1\times 10^{-6}$	& $1.9\times 10^{-6}$
			& $7.0\times 10^{13}$	& $1.0\times 10^{14}$	
			& $8.13$ & $92.1$\\
J1833-1034	& $7.7\times 10^{-7}$	& $3.3\times 10^{-6}$
			& $6.3\times 10^{12}$	& $7.2\times 10^{12}$	
			& $4.86$ & $69.8$\\
J1846-0258	& $3.4\times 10^{-6}$	& $2.2\times 10^{-5}$
			& $9.0\times 10^{13}$	& $9.8\times 10^{13}$	
			& $0.73$ & $65.1$\\
	\noalign{\smallskip}
	\hline
\end{tabular}
\end{table*}

%#################################################################

\begin{figure}[ht!]
	\includegraphics[width=\columnwidth]{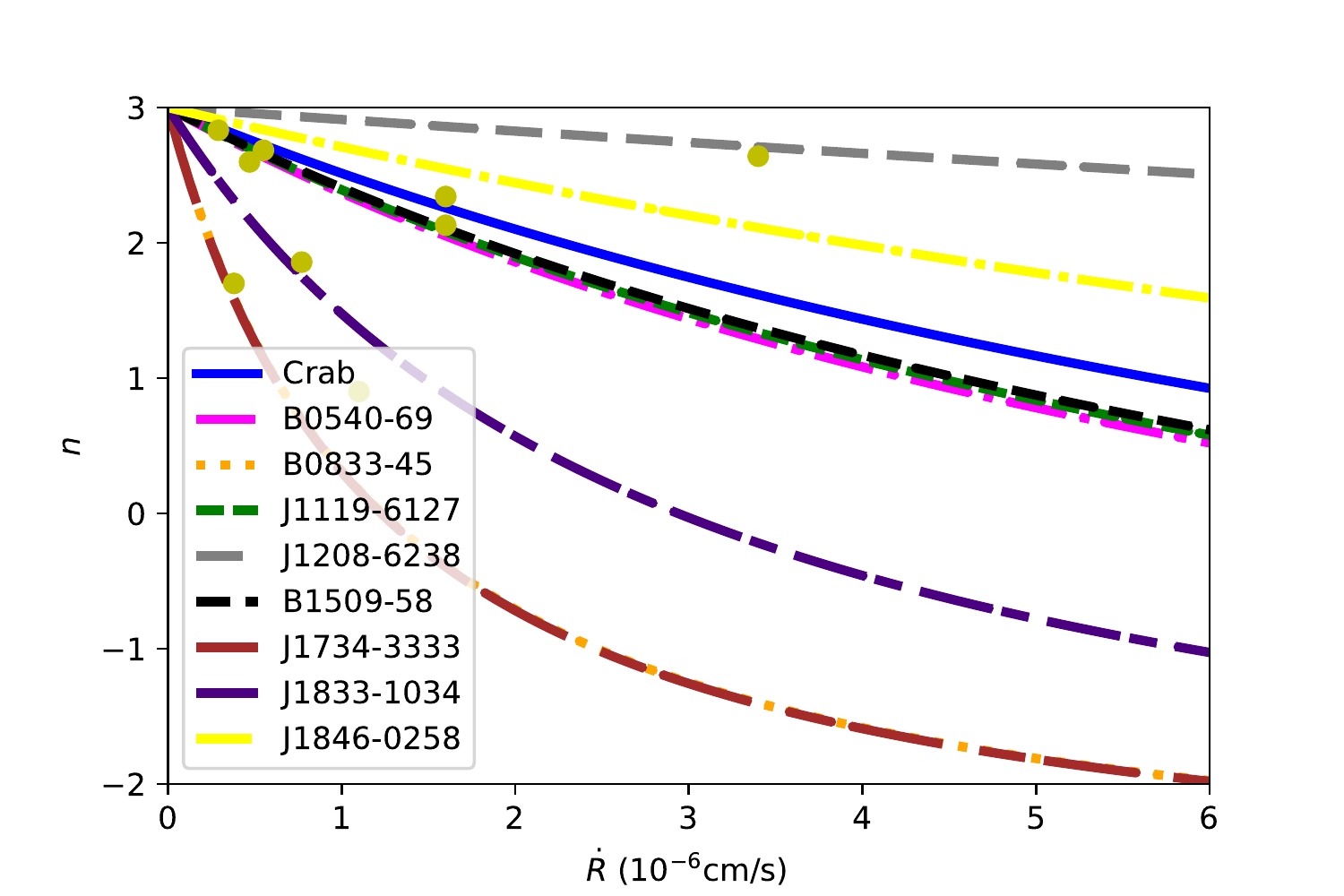}
	\caption{Plots of the braking index, $n$, as a function of the time variation of the displacement parameter, $\dot{R}$ (cm s$^{-1}$), for the pulsars listed in Table \ref{table:01}. The yellow dots indicate observational values for $n$.}
	\label{fig:nVersusDotR}
\end{figure}

Substituting (\ref{eq:Bp2}) in (\ref{eq:braking}) yields
\begin{equation}
n= 3 \frac{\dot{\Omega} / \Omega + \dot{R}/R }{\dot{\Omega}/ \Omega-\dot{R}/R}.
\label{eq:nR}
\end{equation}
In Figure \ref{fig:nVersusDotR} the dependence of the braking index ($n$) on the pulsar core dynamics (quantified by $\dot{R}$) is shown for the pulsars listed in Table \ref{table:01}.

%%%%%%%

%%%%%%%

This equation can be inverted, yielding an expression for the variation in time of the displacement parameter:
\begin{equation}
\dot{R} =  \frac{\dot{\Omega}}{\Omega} \, R \, \frac{n-3}{n+3}.
\label{eq:dotR}
\end{equation}

We used this equation to obtain the values of $\dot{R}$ presented in Table \ref{table:02}, which show that for this sample of pulsars the condition (\ref{eq:condicao}) is not completely fulfilled. The values of the magnetic field given by the canonical model have the same order of magnitude of the values obtained with our model from equation (\ref{eq:Bp2}), as shown in Table \ref{table:02}. Nevertheless, canonical values of the magnetic field should not be used in equations (\ref{eq:braking}) and (\ref{eq:raiovariando}), as they would yield canonical results. The small difference between the values of the magnetic field in the two models is essential to yield observational braking indices.

On the other hand, as a consequence of the magnitude of the terms containing $\dot R$, the magnetic field in equation (\ref{eq:Bp2}), as well as the magnetic dipole moment in equation (\ref{eq:magnMom}) (which varies according to the small fraction $\dot R/R$), have negligible time derivative in our model for practical purposes, as both do in the canonical model. In other words, in our model, despite the existence of a dynamic core that influences the braking index, both the magnetic field and the magnetic dipole moment behave in time as they do in the canonical model. 

The values positive for $\dot{R}$ in Table \ref{table:02} indicate that  those pulsars have displacement parameters that are {\it increasing} with time, i.e., the moments of inertia of the cores of those stars are slowly increasing. As an illustration of such increase we calculated the moment of inertia after the characteristic time, $t_c$, had elapsed for our sample of pulsars (I$_c$ in Table \ref{table:02}), where \cite{Lorimer05}
\begin{equation}
t_c \equiv - \frac{\Omega}{2\dot\Omega}.\label{age}
\end{equation}
This time is normally used as an estimate of the pulsar's age (the {\it spin down 
age}\footnote{The spin down age is obtained in the canonical model under the assumption that pulsars are born with very high angular velocities that} decay in time ($\dot{\Omega} < 0$), resulting in $t_c > 0$. Here we adopt $t_c$ as a definition.), 
and it indicates that Vela (B0833-45) as well as J1734-3333 are the oldest of the sample (see Table \ref{table:02}). 

According to our model and assuming that all pulsars from Table \ref{table:01} have had  $ R = 10 $ km since they were born, after their characteristic times have elapsed all of them would have moments of inertia $I_{char}$ between 60 and 90 M$_\odot$ km$^2$ (see Table \ref{table:02}). These are values within ranges obtained in theoretical models \cite{latt12}. Notably, the  increase in the moment of inertia of B1509-58  during its characteristic time relative to the moment of inertia at birth ($I_0 =$ 56 M$_\odot$ km$^2$) is the smallest in the sample and yet it is relevant for the braking index physics of this model.

Using the definition (\ref{age}) for the characteristic time in the equation for the braking index, (\ref{eq:nR}), we obtain
\begin{equation}
n = 3 \frac{R-2\, \dot{R} \, t_c}{R + 2\, \dot{R} \, t_c}.
\label{eq:n_tc}
\end{equation}

This expression shows that, for a given core dynamics (quantified by a fixed $\dot{R}$), older pulsars (where ``old'' refers to high $t_c$) are expected to have lower braking indices. This is indeed the case shown in Figure \ref{fig:nVersusDotR}, where the curve at the top corresponds to the youngest pulsar of the sample, while the curve at the bottom corresponds to the oldest one.

 For an increasing displacement parameter a pulsar's moment of inertia should increase as well. We analyze this variation noting, in Table \ref{table:02},  that the displacement parameter varies with time within the range  $2.9\times 10^{-7}$ cm s$^{-1}$ to $3.4\times 10^{-6}$ cm s$^{-1}$. The corresponding moments of inertia after the characteristic time has elapsed (I$_c$) lie within the range [57.6, 92.1] M$_{\odot}$ km$^2$. Taking as a reference Figure 2 of Ref. \cite{lat07}, which displays star mass versus radius for a number of equations of state, for spherical neutron stars moments of inertia would lie approximately between 35 and 180 M$_\odot$ km$^2$. Therefore, the values of I$_c$ are well within this theoretical range. 

According to our model the variation in moment of inertia during a time interval $\Delta t$ due to an  increase in the displacement parameter is, to leading order in $\dot R$, given by $\Delta$I = 2 $ \Delta t \dot R I_0$/R. Such changes are small for typical observational time intervals; for the pulsars in Table \ref{table:02}  after a year $\Delta$I would be of the order of 10$^{-3}$ M$_\odot$ km$^2$.

Regarding the centrifugal force, $ F_{cf}= I\Omega ^2 / R$, it decreases in time in the canonical model because so does $\Omega$. We now  show that in our model it is also expected to decrease despite an increasing moment of inertia. Indeed, considering a time-varying displacement parameter given by eqn. (\ref{eq:rft1}) the time derivative of this force is
\[
 \dot{F}_{cf}= \frac{ I \Omega ^2}{R} \left( 2\frac{\dot{\Omega}}{\Omega} + \frac{\dot I}{I} \right) = \frac{ 2I \Omega ^2}{R^2} \left( \frac{R\dot{\Omega}}{\Omega} +\dot R \right). 
\]
This derivative is negative in our model because the first term in the parenthesis is negative and, in modulus, is larger than the second one (see Table \ref{table:02}). In other words, physically the centrifugal force decays because the decrease in the pulsar's angular frequency is faster (relative to the angular frequency involved)  than the simultaneous relative increase in moment of inertia.

%#################################################################

%--------------------------------------------------------------------

\section{Interpreting the Changing Moment of Inertia}\label{sec:4}

By hypothesizing that the moment of inertia could change in time due to motions of the superfluid neutron core, we found that such change could account for the observed braking indices if the pulsar's moment of inertia was increasing due to processes in the core that could be mathematically described by the increase in a displacement parameter. The change in this parameter would be so slow ($\dot R$ is 1.4 cm/day for the Crab pulsar) that even in a relatively long time interval (thousands of years)  the moment of inertia would not depart from values obtained in theoretical models and the centrifugal force would be negative. 

The change in the displacement parameter can be considered a mathematical tool to quantify the complex variations in mass distribution of the superfluid core in only one parameter, namely $\dot{R}$. Indeed, in equation (\ref{eq:314}), the radiated power, quantified in the RHS, originates from a variation of the star crust's angular frequency (first term in the LHS) {\it and} a variation in moment of inertia (second term in the RHS). The variation of the star crust's angular frequency is also present in the canonical model and does not suffice to explain the observed braking indices. The variation in the moment of inertia, introduced in our model, therefore must account for all other physical characteristics of the system that influence the braking index related to mass displacements  in the star's inner structure.

The very slow increase of the core's moment of inertia suggests a gradual accommodation of stellar structures to the continuous change in angular velocity. This is likely to reflect in the crust, where stresses could gradually build up in some regions due to inhomogeneities and eventually ruptures could occur there, perhaps in the form of starquakes. Particularly strong quakes could even disrupt the rotational motion, allowing their observation in the form of glitches in the pulsed radiation.

In order to inspect the meaning of a time-varying displacement parameter, we focus on the conventional model for a pulsar as a spinning neutron star with a thin crust and a large core. In the latter, superconducting protons are assumed to coexist with a large amount of superfluid neutrons together with a smaller amount of relativistic degenerate electrons \cite{rud2006,rud2009,ho12,sou16}.

It is noteworthy that parallel to the pulsar's spin axis the rotating superfluid is expected to form a practically uniform array of co-rotating quantized Onsager-Feynman vortex lines, whose distribution and {\it motion} govern the star's rotational dynamics \cite{onsager49,feynman55,gug2016}. As we presented in Section \ref{sec:3:A}, these lines must expand (move outward) when the neutron star spins down \cite{mus1985,sauls89,sri1990,cham2008}, meaning that a vortex at a distance r$_{\bot}$ from the spin axis would move outward until r$_{\bot}$ reached the core's neutron superfluid radius, which is approximately the radius of the star since the crust is thin. As a consequence of the law of vortex conservation \cite{ghosh07}, while the neutron star spins down the superfluid core responds by destroying the vortices in proportion: the radial outward flow of the vortices yields the annihilation of the required number at the interface between the inner crust and the superfluid core. Such change in the number of vortices implies changes in mass distribution in the core, yielding changes the core's moment of inertia.

For spinning-down pulsars cold enough to have superfluid cores vortex outward velocities are generally less than the cm/day \cite{rud2006,rud2009}, as we calculated in Section \ref{sec:3:A}. The coincidence between this estimate and the order of the increase in the displacement parameter in our model is evident and in fact has the following explanation.

Assuming that angular momentum L = I $\Omega$ is conserved, then $\dot L = \dot I \Omega + I \dot \Omega = 0$. As the moment of inertia of the core (I$_n$) is much larger than the crust's (I$_c$) and the former changes with time while the latter does not, then angular momentum conservation yields 
\[
\dot I_n \Omega = - I_n \dot \Omega. 
\]
Applying the expression for $I_n$, eqn. (\ref{eq:momInCore}), in the above equation results in
\[
\lambda M_n R(t) 2\dot{R} \Omega = - \lambda M_n R^2(t) \dot \Omega
\]
which, for observational time intervals, can be written as
\begin{equation}
\frac{\dot \Omega}{\Omega} = -\frac{2 \dot R}{R_n}.
\label{cap:04:eq:37_nossa}
\end{equation}

This equation is almost identical to eqn. (\ref{cap:04:eq:37_b}). They become identical if we identify
\begin{equation}
  \dot R = v_R.
\label{dRigualvR}
\end{equation}

We now note that in our model $\dot R$ is given by eqn. (\ref{eq:dotR}).  This expression only assumes the form of eqn. (\ref{cap:04:eq:37_nossa}) when n=1. Therefore, only pulsars with braking indices close to 1 would allow the identification (\ref{dRigualvR}). This is the case for J1734-3333, and the values in columns 2 and 3 of Table \ref{table:02} indicate such identification. Therefore, when n=1 we explain this value for the braking index by the changes in moment of inertia due to mass motions inside the core that occur because of the outward motion of superfluid vortex lines. 

For young, isolated pulsars that do not present angular momentum conservation (n $\neq$ 1) we compare eqn. (\ref{eq:dotR}) to eqn. (\ref{cap:04:eq:37_b}) with $R = R_0 \approx R_n$ and identify:
\begin{equation}
  \dot R =  2  \left(\frac{3-n}{3+n}\right) v_R .
\label{dRquasevR}
\end{equation}
In such cases the moment of inertia of the pulsar is changing by a fraction of the contribution due to vortex line motions, indicating that other kinds of mass motions can be occurring in the superfluid core, perhaps linked to the lack of conservation of the angular momentum.

The presence of $\dot{R}$ in the expression for the magnetic field, eqn. (\ref{eq:Bp2}), may inform about how the dynamic of the star core could influence that field. The application of expression (\ref{eq:dotR}) to eqn. (\ref{eq:Bp2}) indicates that the smaller the braking index the stronger the influence of internal processes on the magnetic field.

As observational data have allowed us to obtain numerical values for the parameter $\dot{R}$, we will use this information in the next section to make a few predictions.

%----------------------------------------------------------
\section{Predicting pulsars' braking indices} \label{sec:5}

Assuming that our sample of pulsars is representative of young, isolated pulsars, we used the lowest and highest values of $\dot{R}$  in Table \ref{table:02} to predict ranges of braking indices of other young pulsars. This means that we calculated the lowest range value applying $\dot{R} = 2.9 \times  10^{-7}$ cm s$^{-1}$ to equation (\ref{eq:n_tc}), while the highest one was found applying $\dot{R} = 3.4 \times  10^{-6}$ cm s$^{-1}$ to that equation.

To illustrate this procedure we used the pulsars whose data are listed in Table \ref{table:03}. They were chosen because either their observational braking indices, $n_{obs}$, were published in the literature or they were theoretically investigated elsewhere \cite{magalhaes12}. 

The range that our model predicts for their braking indices is consistent with our previous considerations. The high braking index reported for J1640-4631 by Ref. \cite{archibald16} is yet to be confirmed since those authors could not rule out contamination due to an unseen glitch recovery; our model predicts a more conventional value of $n$ for it. If pulsars with such braking indices are confirmed to exist we will have to analyze their physical characteristics in order to properly explain their n-values using our model. For example, it is known that for indices within the range $3 < n < 5$ there is also loss of energy by emission of gravitational waves (de Araujo, Coelho e Costa 2016). The ranges predicted for pulsars J1124-5916 and J1418-6058 are significantly coincident with the ranges predicted in Ref. \cite{magalhaes12}.

%#################################################################
\begin{table*}{}
	\caption{\label{table:03} Angular velocity ($\Omega$) and its first time derivatives for  other young pulsars. The braking index values, $n_{obs}$, were published. The spin down age is $t_c$, while the $n$ range presented in the last column is the prediction of our model.}
	\centering
        \small{
	\begin{tabular}{cccccccc}
		\hline
		\noalign{\smallskip}
Pulsar & $\Omega$  & $\dot{\Omega}$ ($\times 10^{-10}$) & $\ddot{\Omega}$ ($\times 10^{-21}$)  & Refs. & $n_{obs}$ & $t_c$ & $n$ range \\
& rad s$^{-1}$  & rad s$^{-2}$ & rad s$^{-3}$  & & & ys. & (this model)\\
\noalign{\smallskip}
\hline
\noalign{\smallskip}
	J1640-4631 & 30.4320477075	& -1.433053	& 2.12		& \cite{archibald16}	& 3.15	& 3367	& [0.5, 2.7] \\
	J1124-5916 & 46.377			& -2.5762	& -5.4034	& \cite{ray11}			& -		& 2854	& [0.7, 2.7] \\
	J1418-6058 & 56.822			& -0.8705	& 4.0211	& \cite{kramer03}		& -		& 10349	& [-1.1, 2.1] \\
	B1800-21	& 47.014		& -0.4730	& 0.088		& \cite{espinoza16}		& 1.9	& 15800 & [-1.6, 1.7]\\  
	B1823-13	& 61.920		& -0.4595	& 0.069		& \cite{espinoza16}		& 2.2	& 21400 & [-1.9, 1.3]\\

	   	\noalign{\smallskip}
	   	\hline
	   \end{tabular}}
	\end{table*}
%#################################################################

\section{Conclusions}
\label{sec:conc}

In this work we presented a new model to investigate pulsars' braking indices, which admits a time-varying moment of inertia for the star due to time-varying mass displacements in the superfluid core quantified by $\dot R$. We related the braking index to $\dot{R}$ and found estimates for the value of this parameter were based on observational data. We interpret $\dot{R}$ as a parameter that summarizes different influences on the pulsar's rotation dynamics due to the pulsar's core dynamics. 

We used the values estimated for $\dot{R}$ to predict ranges of values of braking indices for other young pulsars. As more braking indices are observationally determined in the future, the more accurate the statistical knowledge of $\dot{R}$ will be.

We showed that when the braking index is near 1, as is the case for J1734-3333, according to our model the increase in moment of inertia is related to the dynamics of superfluid vortex lines in the pulsar's core due to the coincidence between the estimated value for $\dot{R}$ and the approximate speed of travel of the vortex lines. For pulsars with $n>1$ the motion of such vortex lines seems to have less influence on the value of $n$ as it approaches $n=3$, while pulsars with $n<1$ would present other factors influencing $n$ besides superfluid vortex line motions in the core.

Other investigations may still be conducted in the context of this model, such as: the definition of the family of pulsars for which the range of values obtained for the radius's variation is applicable; verification of the applicability of this model to magnetars and to pulsars with magnetic fields with values different from those of the pulsar sample used here, as well as to pulsars with braking indices greater than 3;  analysis of the model in the presence of other mass values.

%\begin{acknowledgements}

\acknowledgments
HOO thanks \emph{Coordena\c{c}\~{a}o de Aperfei\c{c}oamento de Pessoal de N\'{i}vel Superior} (CAPES) for the financial support. NSM acknowledges the National Institute of Science and Technology in Astrophysics (INCT-A, Brazil; a joint CNPq and FAPESP project, FAPESP grant \# 2008/57807-5) for support. GAC thanks CAPES for financial support under the process PDSE \# 88881.188302/2018-01. CF thanks CPNQ for support, CNPQ grant \# 309098/2017-3. The authors acknowledge FAPESP for support under the thematic project \# 2013/26258-4.
% and thank Manuel M. B. Malheiro de Oliveira for comments and suggestions during the preparation of the manuscript.
%\end{acknowledgements}

\end{document}